\documentclass[twocolumn,english]{IEEEtran}
\usepackage[T1]{fontenc}
\usepackage[latin1]{inputenc}
\usepackage{subfigure}
\usepackage{amsmath}
\usepackage{graphicx}
\usepackage{amssymb}

\makeatletter
  \newtheorem{thm}{Theorem}
  \newtheorem{example}{Example}
  \newtheorem{lemma}{Lemma}
  \newtheorem{prop}{Proposition}
  \newtheorem{remrk}{Remark}

\usepackage{cite}
\usepackage{url}
\usepackage{bm}
\usepackage{bbm}

\usepackage{babel}
\makeatother
\begin{document}

\title{How Many Users should be Turned On\\
in a Multi-Antenna Broadcast Channel?$^{*}$}

\author{Wei Dai, Youjian (Eugene) Liu and Brian Rider\\
 University of Colorado at Boulder\\
425 UCB, Boulder, CO, 80309, USA\\
wei.dai@colorado.edu, eugeneliu@ieee.org, brian.rider@colorado.edu}

\maketitle

\begin{abstract}
This paper considers broadcast channels with $L$ antennas at the
base station and $m$ single-antenna users, where each user has perfect
channel knowledge and the base station obtains channel information
through a finite rate feedback. The key observation of this paper
is that the optimal number of on-users (users turned on), say $s$,
is a function of signal-to-noise ratio (SNR) and other system parameters.
Towards this observation, we use asymptotic analysis to guide the
design of feedback and transmission strategies. As $L$, $m$ and
the feedback rates approach infinity linearly, we derive the asymptotic
optimal feedback strategy and a realistic criterion to decide which
users should be turned on. Define the corresponding asymptotic throughput
per antenna as the \emph{spatial efficiency}. It is a function of
the number of on-users $s$, and therefore, $s$ should be appropriately
chosen. Based on the above asymptotic results, we also develop a scheme
for a system with finite many antennas and users. Compared with other
works where $s$ is presumed constant, our scheme achieves a significant
gain by choosing the appropriate $s$. Furthermore, our analysis and
scheme is valid for heterogeneous systems where different users may
have different path loss coefficients and feedback rates.
\end{abstract}
\begin{keywords}
broadcast channel, finite rate feedback, spatial efficiency
\end{keywords}
\renewcommand{\thefootnote}{\fnsymbol{footnote}} \footnotetext[1]{This work is supported by NSF Grant DMS-0505680 and Thomson Inc.} \renewcommand{\thefootnote}{\arabic{footnote}} \setcounter{footnote}{0}

\section{\label{sec:Introduction}Introduction}

It is well known that multiple antennas can improve the spectral efficiency.
This paper considers broadcast channels with $L$ antennas at the
base station and $m$ single-antenna users. To achieve the full benefit,
perfect channel state information (CSI) is required at both receiver
and transmitter. Perfect CSI at the receiver can be obtained by estimation
from the received signal. However, if CSI at the transmitter (CSIT)
is obtained from feedback, perfect CSIT requires an infinite feedback
rate. As this is not feasible in practice, it is important to analyze
the effect of finite rate feedback and design efficient strategy accordingly.

The feedback models for broadcast channels are described as follows.
To save feedback rate on power control, we assume a power on/off strategy
where each user is either turned on with a constant power or turned
off, and that the number of on-users (the users turned on) is a constant,
say $s$, independent of the channel realization. For any given channel
realization, the users quantize their channel states into finite bits
and feedback the corresponding indices to the base station. After
receiving the feedback from users, the base station decides which
users should be turned on and then forms beamforming vectors for transmission. 

Broadcast channels with feedback have been studied in \cite{Sharif_IT05_MIMO_BC_Feedback,Jindal_IT06sub_BC_Feedback}.
Ideally, if the base station has the perfect CSI, zero-forcing transmission
avoids interference among users. However, with only finite rate feedback
on CSI, the base station does not know the perfect channel state information
and therefore interference from other users is inevitable. The interference
gets so strong at high signal-to-noise ratio (SNR) regions that the
system throughput is upper bounded by a constant even when SNR approach
infinity. This phenomenon is called interference domination and was
reported on in \cite{Sharif_IT05_MIMO_BC_Feedback,Jindal_IT06sub_BC_Feedback}.
The analysis is based on the assumption that the number of on-users
$s$ always equals to the number of antennas at the base station $L$
($L\le m$ is typically assumed). To limit the interference to a desired
level, Sharif and Hassibi let $m$ grow exponentially with $L$ such
that there are $L$ near orthogonal users with high probability \cite{Sharif_IT05_MIMO_BC_Feedback}.
In both scenarios, a homogeneous system is assumed where all the users
share the same path loss coefficient and feedback resource. 

Different from the above approaches, this paper studies heterogeneous
broadcast systems, where different users may have different path loss
coefficients and feedback rates. Furthermore, different from \cite{Sharif_IT05_MIMO_BC_Feedback},
we focus on systems with a relatively small number of users. Note
that a cooperative communication network can often be viewed as a
composition of multi-access and broadcast sub-systems with a small
number of users. Research on broadcast systems of small size provides
insights into cooperative communications. 

For such systems, we solve the interference domination problem by
choosing the appropriate number of on-users $s$. The reason that
random beams construction in \cite{Sharif_IT05_MIMO_BC_Feedback}
fails in our small size systems is elaborated in Theorem \ref{thm:Random-Beams}.

Our solution is based on the asymptotic analysis where $L,m,s$ and
the feedback rates approach infinity linearly with constant ratios
among them. This type of asymptotics is applied to systems of small
size. The main asymptotic results are: 

\begin{itemize}
\item It is asymptotically optimal to quantize the channel directions only
and ignore the channel magnitude information. The asymptotically optimal
feedback function and codebook are derived accordingly.
\item A realistic on/off criterion is proposed to decide which users should
be turned on.
\item The corresponding throughput per antenna converges to a constant,
defined as the \emph{spatial efficiency}. It is a function of the
normalized number of on-users $\bar{s}=\frac{s}{L}$. Further, there
exists a unique $\bar{s}\in\left(0,1\right)$ to maximize the the
spatial efficiency. 
\end{itemize}

We develop a scheme to choose the appropriate $s$ for systems with
finite $L$ and $m$. Simulations show that the gain achieved by choosing
$s$ is significant compared with the strategies where $s\equiv L$.
In addition, our scheme has the following advantages.

\begin{itemize}
\item It is valid for heterogeneous systems. 
\item The set of on-users is independent of the channel realization. As
a result, computation complexity is low since we do not have to perform
a user selection computation every fading block.
\item Only on-users need to feedback CSI, which saves a large amount of
feedback resource. 
\end{itemize}

This paper is organized as follows. The system model is introduced
in Section \ref{sec:System-Model}. Then Section \ref{sec:Asymptotic-Analysis}
performs the asymptotic analysis obtaining insights into system design,
and quantifies the spatial efficiency. Based on the asymptotic results,
a practical scheme is developed in Section \ref{sec:Finite-System-Design}
for systems with finite many antennas and users. Finally, conclusions
are summarized in Section \ref{sec:Conclusion}.

\section{\label{sec:System-Model}System Model}

Consider a broadcast channel with $L$ antennas at the base station
and $m$ single-antenna users. Assume that the base station employs
zero forcing transmitter. Let $\gamma_{i}\in\mathbb{R}\backslash\mathbb{R}^{-}$
($1\le i\le m$) be the path loss coefficient for user $i$. Then
the signal model for user $i$ is\[
Y_{i}=\sqrt{\gamma_{i}}\mathbf{h}_{i}^{\dagger}\left(\sum_{j=1}^{m}\mathbf{q}_{j}X_{j}\right)+W_{i},\]
where $Y_{i}\in\mathbb{C}$ is the received signal for user $i$,
$\mathbf{h}_{i}\in\mathbb{C}^{L\times1}$ is the channel state vector
for user, $\mathbf{q}_{j}\in\mathbb{C}^{L\times1}$ is the zero-forcing
beamforming vector for user $j$, $X_{j}\in\mathbb{C}$ is the source
signal for the user $j$ and $W_{i}\in\mathbb{C}$ is the complex
Gaussian noise with zero mean and unit variance $\mathcal{CN}\left(0,1\right)$.
Here, we assume that $\mathbf{q}_{j}^{\dagger}\mathbf{q}_{j}=1$ and
the Rayleigh block fading channel model: the entries of $\mathbf{h}_{i}$
are independent and identically distributed (i.i.d.) $\mathcal{CN}\left(0,1\right)$.
Without loss of generality, we assume that $L\le m$; if $L>m$, adding
$L-m$ users with $\gamma_{i}=0$ yields an equivalent system with
$L^{\prime}=m$. 

For the above broadcast system, it is natural to assume a total power
constraint $\sum_{i=1}^{m}\mathrm{E}\left[\left|X_{i}\right|^{2}\right]\leq\rho$.
Further, for implementation simplicity, we assume a power on/off strategy
with a constant number of on-users as follows. 

\begin{description}
\item [{A1)}] Power on/off strategy: a source $X_{i}$ is either turned
on with a constant power $P_{\mathrm{on}}$ or turned off. It is motivated
by the fact that this strategy is near optimal for single user MIMO
system \cite{Dai_05_Power_onoff_strategy_design_finite_rate_feedback}.
\item [{A2)}] A constant number of on-users: we assume that the number
of on-users $s$ ($1\le s\le m$) is a constant independent of the
specific channel realizations. With this assumption, $P_{\mathrm{on}}=\frac{\rho}{s}$.
Here, $s$ is a function of $\rho$, $\gamma_{i}$ and feedback rate.
This assumption is different from the one in \cite{Sharif_IT05_MIMO_BC_Feedback,Jindal_IT06sub_BC_Feedback},
where $s=L$ always ($L\leq m$ is assumed there). 
\end{description}

The finite rate feedback model is then described as follows. Assume
that both base station and user $i$ knows $\gamma_{i}$%
\footnote{There are many ways in which the base station obtains $\gamma_{i}$.
A simple example could be that the base station measures the feedback
signal strength. %
} but only user $i$ knows the channel state realization $\mathbf{h}_{i}$
perfectly. For given channel realizations $\mathbf{h}_{1}\cdots\mathbf{h}_{m}$,
user $i$ quantizes his channel $\mathbf{h}_{i}$ into $R_{i}$ bits
and then feeds the corresponding index to the base station. Formally,
let $\mathcal{B}_{i}=\left\{ \hat{\mathbf{h}}\in\mathbb{C}^{L\times1}\right\} $
with $\left|\mathcal{B}_{i}\right|=2^{R_{i}}$ be a channel state
codebook for user $i$. Then the quantization function is given by\begin{align*}
\mathfrak{q}:\mathbb{C}^{L\times1} & \rightarrow\mathcal{B}_{i}\\
\mathbf{h}_{i} & \mapsto\hat{\mathbf{h}}_{i}.\end{align*}
In Section \ref{sub:quantization-function} and \ref{sub:codebooks},
we will show how to design $\mathfrak{q}$ and $\mathcal{B}$ respectively.

After receiving feedback information from users, the base station
decides which $s$ users should be turned on and forms zero-forcing
beamforming vectors for them. Let $A_{\mathrm{on}}$ be the set of
$s$ on-users. The zero-forcing beamforming vectors $\mathbf{q}_{i}$'s
$i\in A_{\mathrm{on}}$ is calculated as follows. Let $\mathcal{P}_{i}^{\perp}$
be the plane generated by $\left\{ \hat{\mathbf{h}}_{j}:\; j\in A_{\mathrm{on}}\backslash\left\{ i\right\} \right\} $.
Let $\mathcal{P}_{i}$ be the orthogonal complement of $\mathcal{P}_{i}^{\perp}$
and $t$ be the dimensions of $\mathcal{P}_{i}$. Define $\mathbf{T}_{i}\in\mathbb{C}^{L\times t}$
the matrix whose columns are orthonormal and span the plane $\mathcal{P}$.
Then $\mathbf{q}_{i}$ is the \emph{unitary projection} of $\hat{\mathbf{h}}_{i}$on
$\mathbf{T}_{i}$\begin{equation}
\mathbf{q}_{i}:=\frac{\mathbf{T}_{i}\mathbf{T}_{i}^{\dagger}\hat{\mathbf{h}}_{i}}{\left\Vert \mathbf{T}_{i}\mathbf{T}_{i}^{\dagger}\hat{\mathbf{h}}_{i}\right\Vert }.\label{eq:zero-forcing-rule}\end{equation}
Here, if $s=1$ and $A_{\mathrm{on}}=\left\{ i\right\} $, $\mathbf{T}_{i}$
is a $L\times L$ unitary matrix and $\mathbf{q}_{i}=\hat{\mathbf{h}}_{i}/\left\Vert \hat{\mathbf{h}}_{i}\right\Vert $.

\section{\label{sec:Asymptotic-Analysis}Asymptotic Analysis }

In order to obtain insights into system design, this section performs
asymptotic analysis by letting $L,m,R_{i}\mathrm{'s}\rightarrow\infty$
linearly. The quantization function $\mathfrak{q}$ and asymptotically
optimal codebook $\mathcal{B}_{i}$ are derived in Section \ref{sub:quantization-function}
and \ref{sub:codebooks} respectively. Then Section \ref{sub:On/off-Criterion}
develops a realistic on/off criterion to decide which users should
be turned on. Finally Section \ref{sub:The-Spatial-Efficiency} computes
the corresponding spatial efficiency.

\subsection{\label{sub:quantization-function}Design of Quantization Function}

Generally speaking, full information of $\mathbf{h}_{i}$ contains
the direction information $\mathbf{v}_{i}:=\mathbf{h}_{i}/\left\Vert \mathbf{h}_{i}\right\Vert $
and the magnitude information $\left\Vert \mathbf{h}_{i}\right\Vert $.
In our Rayleigh fading channel model, it is well known that $\mathbf{v}_{i}$
and $\left\Vert \mathbf{h}_{i}\right\Vert $ are independent. Intuitively,
joint quantization of $\mathbf{v}_{i}$ and $\left\Vert \mathbf{h}_{i}\right\Vert $
is preferred.

Interestingly, Theorem \ref{thm:magnitude-concentration} implies
that there is no need to quantize the channel magnitudes. Indeed,
as $L,m\rightarrow\infty$ linearly, all users' channel magnitudes
concentrate on a single value with probability one.

\begin{thm}
\label{thm:magnitude-concentration}For $\forall\epsilon>0$, as $L,m\rightarrow\infty$
with $\frac{m}{L}\rightarrow\bar{m}\in\mathbb{R}^{+}$,\[
\Pr\left(\underset{1\le i\le m}{\max}\;\frac{1}{L}\left\Vert \mathbf{h}_{i}\right\Vert ^{2}\ge1+\epsilon\right)\rightarrow0,\]
and \[
\Pr\left(\underset{1\le i\le m}{\min}\;\frac{1}{L}\left\Vert \mathbf{h}_{i}\right\Vert ^{2}\le1-\epsilon\right)\rightarrow0.\]

\end{thm}

The proof of Theorem \ref{thm:magnitude-concentration} is omitted
due to the space limitation. An important fact behind the proof is
that whether the users' channel magnitudes concentrate or not depends
on the relationship between $L$ and $m$: this concentration happens
in our asymptotic region where $L$ and $m$ are of the same order. 

To fully understand Theorem \ref{thm:magnitude-concentration}, it
is important to realize that the Law of Large Numbers does not imply
that all users' channel magnitudes will concentrate. According to
the Law of Large Numbers, $\frac{1}{L}\left\Vert \mathbf{h}_{i}\right\Vert \rightarrow1$
almost surely for any \emph{given} $i$. However, if $m$ approaches
infinity exponentially with $L$, there are certain number of users
whose channel magnitudes are larger than others', and therefore it
may be still beneficial to quantize and feedback channel magnitude
information. This phenomenon is illustrated by the following example. 

\begin{example}
(A case where magnitude information is beneficial) Consider a broadcast
channel with $\gamma_{1}=\cdots=\gamma_{m}=1$. As $L,m\rightarrow\infty$
with $\log\left(m\right)/L\rightarrow\bar{m}^{\prime}\in\mathbb{R}^{+}$,
there exists an $\epsilon>0,$ $\delta_{1}>0$ and $\delta_{2}>0$
such that \[
\frac{1}{L}\log\left|\left\{ i:\;\frac{1}{L}\left\Vert \mathbf{h}_{i}\right\Vert ^{2}>1+\frac{\epsilon}{2}\right\} \right|\rightarrow\delta_{1},\]
and \[
\frac{1}{L}\log\left|\left\{ i:\;\frac{1}{L}\left\Vert \mathbf{h}_{i}\right\Vert ^{2}<1-\frac{\epsilon}{2}\right\} \right|\rightarrow\delta_{2}\]
 with probability one. Note that there are a set of users whose channel
magnitudes are $\epsilon$-larger than another set of users. It may
be worth to let the base station know which users have stronger channels.
\end{example}

Theorem \ref{thm:magnitude-concentration} implies that it is sufficient
to quantize the channel direction information only and omit the channel
magnitude information. For this quantization, the codebook is given
by $\mathcal{B}_{i}=\left\{ \mathbf{p}\in\mathbb{C}^{L\times1}:\;\left\Vert \mathbf{p}\right\Vert =1\right\} $
with $\left|\mathcal{B}_{i}\right|=2^{R_{i}}$. We use the following
quantization function \begin{align}
\mathfrak{q}:\mathbb{C}^{L\times1} & \rightarrow\mathcal{B}_{i}\nonumber \\
\mathbf{h}_{i} & \mapsto\mathbf{p}_{i}=\underset{\mathbf{p}\in\mathcal{B}_{i}}{\arg\;\max}\;\left|\mathbf{v}_{i}^{\dagger}\mathbf{p}\right|,\label{eq:quantization-fn}\end{align}
where $\mathbf{v}_{i}$ is the channel direction vector.

\subsection{\label{sub:codebooks}Asymptotically Optimal Codebooks}

Given the quantization function (\ref{eq:quantization-fn}), the distortion
of a given codebook $\mathcal{B}_{i}$ is the average chordal distance
between the actual and quantized channel directions corresponding
to the codebook $\mathcal{B}_{i}$ and defined as \[
D\left(\mathcal{B}_{i}\right):=1-\mathrm{E}_{\mathbf{h}_{i}}\left[\underset{\mathbf{p}\in\mathcal{B}_{i}}{\max}\left|\mathbf{v}_{i}^{\dagger}\mathbf{p}\right|^{2}\right].\]

The following lemma bounds the minimum achievable distortion for a
given codebook rate (usually called the distortion rate function). 

\begin{lemma}
\label{lem:dist-rate-fn}Define $D^{*}\left(R\right)\triangleq\underset{\mathcal{B}:\;\left|\mathcal{B}\right|\leq2^{R}}{\inf}D\left(\mathcal{B}\right)$.
Then \begin{eqnarray}
 &  & \frac{L-1}{L}2^{-\frac{R}{L-1}}\left(1+o\left(1\right)\right)\leq D^{*}\left(R\right)\nonumber \\
 &  & \quad\quad\quad\leq\frac{\Gamma\left(\frac{1}{L-1}\right)}{L-1}2^{-\frac{R}{L-1}}\left(1+o\left(1\right)\right),\label{eq:quantization-bds}\end{eqnarray}
and as $L$ and $R$ approach infinity with $\frac{R}{L}\rightarrow\bar{r}\in\mathbb{R}^{+}$,
\[
\underset{\left(L,R\right)\rightarrow\infty}{\lim}D^{*}\left(R\right)=2^{-\bar{r}}.\]

\end{lemma}

The following Lemma shows that a random codebook is asymptotically
optimal with probability one.

\begin{lemma}
\label{lem:random-codes-asymptotically-optimal}Let $\mathcal{B}_{\mathrm{rand}}$
be a random codebook where the vectors $\mathbf{p}\in\mathcal{B}_{\mathrm{rand}}$'s
are independently generated from the isotropic distribution. Let $R=\log\left|\mathcal{B}_{\mathrm{rand}}\right|$.
As $L,R\rightarrow\infty$ with $\frac{R}{L}\rightarrow\bar{r}\in\mathbb{R}^{+}$,
for $\forall\epsilon>0$, \[
\underset{\left(L,R\right)\rightarrow\infty}{\lim}\;\mathrm{Pr}\left\{ \mathcal{B}_{\mathrm{rand}}:\; D\left(\mathcal{B}_{\mathrm{rand}}\right)>2^{-\bar{r}}+\epsilon\right\} =0.\]

\end{lemma}

The proofs of Lemma \ref{lem:dist-rate-fn} and \ref{lem:random-codes-asymptotically-optimal}
are given in our paper \cite{Dai_05_Quantization_Grassmannian_manifold}.

Due to the asymptotic optimality of random codebooks, we assume that
the codebooks $\mathcal{B}_{i}$'s $i=1,\cdots,m$ are independent
and randomly constructed throughout this paper.

\subsection{\label{sub:On/off-Criterion}On/off Criterion}

After receiving feedback from users, the base station should decide
which $s$ users should be turned on. 

Ideally, for given channel realizations $\mathbf{h}_{1},\cdots,\mathbf{h}_{m}$,
the optimal set of on users $A_{\mathrm{on}}^{*}$ should be chosen
to maximize the instantaneous mutual information. Note that the base
station only knows the quantized version of channel states $\mathbf{p}_{1},\cdots,\mathbf{p}_{m}$.
It can only estimate the instantaneous mutual information through
$\mathbf{p}_{i}$'s. The set $A_{\mathrm{on}}^{*}$ is given by\begin{align}
A_{\mathrm{on}}^{*}= & \underset{A_{\mathrm{on}}:\;\left|A_{\mathrm{on}}\right|=s}{\arg\;\max}\;\sum_{i\in A_{\mathrm{on}}}\log\left(1+\right.\nonumber \\
 & \quad\quad\quad\left.\frac{\gamma_{i}\frac{\rho}{s}\left|\mathbf{p}_{i}^{\dagger}\mathbf{q}_{i}\right|^{2}}{1+\gamma_{i}\frac{\rho}{s}\sum_{j\in A_{\mathrm{on}}\backslash\left\{ i\right\} }\left|\mathbf{p}_{i}^{\dagger}\mathbf{q}_{j}\right|^{2}}\right).\label{eq:optimal-Aon}\end{align}
However, finding $A_{\mathrm{on}}^{*}$ requires exhaustive search,
whose complexity exponentially increases with $m$. 

The random orthonormal beams construction method in \cite{Sharif_IT05_MIMO_BC_Feedback}
does not work for our asymptotically large system either. In \cite{Sharif_IT05_MIMO_BC_Feedback},
the base station randomly constructs $L$ orthonormal beams $\mathbf{b}_{1},\cdots,\mathbf{b}_{L}$,
finds the users with highest signal-to-noise-plus-interference ratios
(SINRs) through feedback from users, and then transmits to these selected
users. There, the SINR calculation for user $i$ is related to the
quantity $\underset{1\le k\le L}{\max}\left|\mathbf{h}_{i}^{\dagger}\mathbf{b}_{k}\right|$.
However, Theorem \ref{thm:Random-Beams} below shows that in the asymptotic
region where $L$ and $m$ are of the same order, all users' channels
are near orthogonal to all of the $L$ orthonormal beams $\mathbf{b}_{i}$'s.
Therefore, all users' maximum SINRs (maximum over $L$ given orthonormal
beams) approach zero uniformly with probability one. The method in
\cite{Sharif_IT05_MIMO_BC_Feedback} fails  in our asymptotically
large system.

\begin{thm}
\label{thm:Random-Beams}Given $\forall\epsilon>0$ and any $L$ orthonormal
beams $\mathbf{b}_{k}\in\mathbb{C}^{L\times1}$ $1\le k\le L$, as
$L,m\rightarrow\infty$ linearly with $\frac{m}{L}\rightarrow\bar{m}\in\mathbb{R}^{+}$,
\[
\underset{\left(L,m\right)\rightarrow\infty}{\lim}\;\Pr\left(\underset{1\le i\le m,\;1\le k\le L}{\max}\;\frac{1}{L}\left|\mathbf{h}_{i}^{\dagger}\mathbf{b}_{k}\right|>\epsilon\right)=0.\]

\end{thm}

The proof is omitted due to the space limitation.

In this paper, we take another approach where the on/off decision
is independent of channel directions. We start with the throughput
analysis for a specific on-user $i\in A_{\mathrm{on}}$. Note that
\[
Y_{i}=\sqrt{\gamma_{i}}\mathbf{h}_{i}^{\dagger}\mathbf{q}_{i}X_{i}+\left(\sqrt{\gamma_{i}}\mathbf{h}_{i}^{\dagger}\sum_{j\in A_{\mathrm{on}}\backslash\left\{ i\right\} }\mathbf{q}_{j}X_{j}+W\right).\]
The signal power and interference power for user $i$ are given by
\begin{equation}
P_{\mathrm{sig},i}=\frac{\rho}{s}\gamma_{i}\left|\mathbf{h}_{i}^{\dagger}\mathbf{q}_{i}\right|^{2}\label{eq:signal-power}\end{equation}
 and \begin{equation}
P_{\mathrm{int},i}=\frac{\rho}{s}\gamma_{i}\sum_{j\in A_{\mathrm{on}}\backslash\left\{ i\right\} }\left|\mathbf{h}_{i}^{\dagger}\mathbf{q}_{j}\right|^{2}\label{eq:interference-power}\end{equation}
respectively. Note that the influence of the users in $A_{\mathrm{on}}\backslash\left\{ i\right\} $
on user $i$ only occurs through their directions $\mathbf{q}_{j}$'s
$j\in A_{\mathrm{on}}\backslash\left\{ i\right\} $. If the choice
of on-users $A_{\mathrm{on}}$ is independent of their channel directions,
then $\mathbf{h}_{i}$ and $\mathbf{q}_{j}$'s are independent. In
this case, $P_{\mathrm{sig},i}$ and $P_{\mathrm{int},i}$ can be
quantified as $L,m,s,R_{i}\mathrm{'s}\rightarrow\infty$. The result
is given in the following proposition.

\begin{prop}
\label{pro:on-user-i-throughput}Let $\left|A_{\mathrm{on}}\right|=s$
and $L,m,s,R_{i}\mathrm{'s}\rightarrow\infty$ with $\frac{m}{L}\rightarrow\bar{m}$,
$\frac{s}{L}\rightarrow\bar{s}$ and $\frac{R_{i}}{L}\rightarrow\bar{r}_{i}$.
Assume that $\mathbf{v}_{i}$'s $i\in A_{\mathrm{on}}$ are independent.
Then for $\forall i\in A_{\mathrm{on}}$, \[
P_{\mathrm{sig},i}\rightarrow\frac{\rho}{\bar{s}}\gamma_{i}\left(1-2^{-\bar{r}_{i}}\right)\left(1-\bar{s}\right),\]
\[
P_{\mathrm{int},i}\rightarrow\rho\gamma_{i}2^{-\bar{r}_{i}},\]
and therefore\begin{align*}
\mathcal{I}_{i} & :=\log\left(1+\frac{P_{\mathrm{sig},i}}{1+P_{\mathrm{int},i}}\right)\\
 & \rightarrow\log\left(1+\eta_{i}\frac{1-\bar{s}}{\bar{s}}\right),\end{align*}
with probability one, where\begin{equation}
\eta_{i}:=\frac{\rho\gamma_{i}\left(1-2^{-\bar{r}_{i}}\right)}{1+\rho\gamma_{i}2^{-\bar{r}_{i}}}.\label{eq:eta-zf}\end{equation}

\end{prop}

\begin{remrk}
This proposition may not be true if $\mathbf{v}_{j}$'s ($j\in A_{\mathrm{on}}\backslash\left\{ i\right\} $)
are not independent of $\mathbf{v}_{i}$. Indeed, for example, if
other users are chosen such that their channel directions are as orthogonal
to user $i$ as possible, the interference to user $i$ is less than
that achieved by our choice where channel directions are not taken
into consideration. This claim is verified by the fact that $\exists\epsilon>0$
such that \[
\underset{A_{\mathrm{on}}\backslash\left\{ i\right\} }{\min}\;\sum_{j\in A_{\mathrm{on}}\backslash\left\{ i\right\} }\left|\mathbf{h}_{i}^{\dagger}\mathbf{q}_{j}\right|^{2}<\sum_{j\in A_{\mathrm{on},\mathrm{rand}}\backslash\left\{ i\right\} }\left|\mathbf{h}_{i}^{\dagger}\mathbf{q}_{j}\right|^{2}-\epsilon\]
with probability one as $L,m,s\rightarrow\infty$ linearly, where
$A_{\mathrm{on,}\mathrm{rand}}\backslash\left\{ i\right\} $ denotes
a random choice of $A_{\mathrm{on}}\backslash\left\{ i\right\} $. 
\end{remrk}

\begin{remrk}
Proposition \ref{pro:on-user-i-throughput} shows that the user $i$'s
asymptotic throughput is a constant independent of the specific channel
realization $\mathbf{h}_{i}$ with probability one. 
\end{remrk}

Based on Proposition \ref{pro:on-user-i-throughput}, we select the
set of $s$ on-users $A_{\mathrm{on}}$ such that $\left|A_{\mathrm{on}}\right|=s$
and \begin{equation}
A_{\mathrm{on}}=\left\{ i:\;\eta_{i}\ge\eta_{j}\;\mathrm{for}\;\forall j\notin A_{\mathrm{on}}\right\} ;\label{eq:on-off-criterion}\end{equation}
if there are multiple candidates, we randomly choose one of them.
It is the asymptotically optimal on/off selection if the on/off decision
is independent of the channel direction information. The difference
between the throughput achieved by optimal on/off criterion in (\ref{eq:on-off-criterion})
and the proposed one in (\ref{eq:optimal-Aon}) remains unknown.

\subsection{\label{sub:The-Spatial-Efficiency}The Spatial Efficiency}

We define the spatial efficiency (bits/sec/Hz/antenna) as\[
\bar{\mathcal{I}}\left(\bar{s}\right):=\underset{\left(L,m,s,R_{i}\mathrm{'s}\right)\rightarrow\infty}{\lim}\bar{\mathcal{I}}^{\left(L\right)},\]
where $L,m,s,R_{i}\mathrm{'s}\rightarrow\infty$ in the same way as
before, $\bar{\mathcal{I}}^{\left(L\right)}$ is the average throughput
per antenna given by \[
\bar{\mathcal{I}}^{\left(L\right)}:=\mathrm{E}_{\mathcal{B}_{i}\mathrm{'s},\mathbf{h}_{i}\mathrm{'s}}\left[\frac{1}{L}\sum_{i\in A_{\mathrm{on}}}\log\left(1+\frac{P_{\mathrm{sig},i}}{1+P_{\mathrm{int},i}}\right)\right],\]
and $A_{\mathrm{on}}$, $P_{\mathrm{sig},i}$ and $P_{\mathrm{int},i}$
are defined in (\ref{eq:on-off-criterion}), (\ref{eq:signal-power})
and (\ref{eq:interference-power}) respectively. 

We shall quantify $\bar{\mathcal{I}}\left(\bar{s}\right)$ for a given
$\bar{s}$. Define the empirical distribution of $\eta_{i}$ as \[
\mu_{\eta}^{\left(m\right)}\left(\eta\le x\right):=\frac{1}{m}\left|\left\{ \eta_{i}:\;\eta_{i}\le x\right\} \right|,\]
and assume that $\mu_{\eta}:=\lim\mu_{\eta}^{\left(m\right)}$ exists
weakly as $L,m,R_{i}\mathrm{'s}\rightarrow\infty$. In order to cope
with $\mu_{\eta}$'s with mass points, define \[
\int_{x^{+}}^{\infty}f\left(\eta\right)d\mu_{\eta}:=\underset{\Delta x\downarrow0}{\lim}\int_{x+\Delta x}^{\infty}f\left(\eta\right)d\mu_{\eta}\]
for $\forall x\in\mathbb{R}$, where $f$ is a integrable function
with respect to $\mu_{\eta}$. Then $\bar{\mathcal{I}}\left(\bar{s}\right)$
is computed in the following theorem.

\begin{thm}
\label{thm:spatial-efficiency}Let $L,m,s,R_{i}\mathrm{'s}\rightarrow\infty$
with $\frac{m}{L}\rightarrow\bar{m}$, $\frac{s}{L}\rightarrow\bar{s}$
and $\frac{R_{i}}{L}\rightarrow\bar{r}_{i}$. Define \[
\eta_{\bar{s}}:=\sup\left\{ \eta:\;\bar{m}\int_{\eta}^{\infty}d\mu_{\eta}>\bar{s}\right\} .\]
Then as $\bar{s}\notin\left(0,1\right)$, $\bar{\mathcal{I}}\left(\bar{s}\right)=0$.
If $\bar{s}\in\left(0,1\right)$, \begin{align}
\bar{\mathcal{I}}\left(\bar{s}\right) & =\bar{m}\int_{\eta_{\bar{s}}^{+}}^{\infty}\log\left(1+\eta\frac{1-\bar{s}}{\bar{s}}\right)d\mu_{\eta}\nonumber \\
 & \quad+\left(\bar{s}-\bar{m}\int_{\eta_{\bar{s}}^{+}}^{\infty}d\mu_{\eta}\right)\log\left(1+\eta_{\bar{s}}\frac{1-\bar{s}}{\bar{s}}\right).\label{eq:zf-spatial-efficiency}\end{align}

\end{thm}

We are also interested in finding the optimal $\bar{s}$ to maximize
$\bar{\mathcal{I}}\left(\bar{s}\right)$. Unfortunately, $\bar{\mathcal{I}}\left(\bar{s}\right)$
is not a concave function of $\bar{s}$ in general. Furthermore, the
measure $\mu_{\eta}$ may contain mass points. The optimization of
$\bar{\mathcal{I}}\left(\bar{s}\right)$ is therefore a non-convex
and non-smooth optimization problem. The following theorem provides
a criterion to find the optimal $\bar{\mathcal{I}}\left(\bar{s}\right)$. 

\begin{thm}
\label{thm:optimal-s}$\bar{\mathcal{I}}\left(\bar{s}\right)$ is
maximized at a unique $\bar{s}^{*}\in\left(0,1\right)$ such that\begin{align}
 & 0\in\left[\underset{\Delta\bar{s}\rightarrow0}{\lim\;\inf}\frac{\bar{\mathcal{I}}\left(\bar{s}^{*}\right)-\bar{\mathcal{I}}\left(\bar{s}^{*}-\Delta\bar{s}\right)}{\Delta\bar{s}},\right.\nonumber \\
 & \quad\quad\left.\underset{\Delta\bar{s}\rightarrow0}{\lim\;\sup}\frac{\bar{\mathcal{I}}\left(\bar{s}^{*}\right)-\bar{\mathcal{I}}\left(\bar{s}^{*}-\Delta\bar{s}\right)}{\Delta\bar{s}}\right].\label{eq:optimal-s}\end{align}

\end{thm}

The proof is omitted due to the space limitation. The $\bar{\mathcal{I}}\left(\bar{s}^{*}\right)$
is the maximum achievable spatial efficiency for the proposed power
on/off strategy.

\section{\label{sec:Finite-System-Design}Finite Dimensional System Design}

Based on the asymptotic results in Theorem \ref{thm:spatial-efficiency}-\ref{thm:optimal-s},
we now propose a scheme for systems with finite $L$ and $m$.

\subsection{Throughput Estimation for Finite Dimensional Systems}

While asymptotic analysis provide many insights, we do not apply asymptotic
results directly for a finite dimensional system. The reason is that
in asymptotic analysis $\frac{1}{L}\rightarrow0$ while $\frac{1}{L}$
cannot be ignored for a system with small $L$. In the following,
we first calculate the main order term of the throughput for user
$i\in A_{\mathrm{on}}$ and then explain the difference between asymptotic
analysis and finite dimensional system analysis explicitly.

To obtain the main order term, proceed as follows. Note that the throughput
for user $i\in A_{\mathrm{on}}$ ($\left|A_{\mathrm{on}}\right|=s$)
is \begin{align*}
\mathcal{I}_{i} & =\mathrm{E}\left[\log\left(1+\frac{P_{\mathrm{sig},i}}{1+P_{\mathrm{int},i}}\right)\right]\\
 & =\log\left(1+\frac{\mathrm{E}\left[P_{\mathrm{sig},i}\right]}{1+\mathrm{E}\left[P_{\mathrm{int},i}\right]}\right)\\
 & \quad+\mathrm{E}\left[\log\left(\frac{1+P_{\mathrm{sig},i}+P_{\mathrm{int},i}}{1+\mathrm{E}\left[P_{\mathrm{sig},i}\right]+\mathrm{E}\left[P_{\mathrm{int},i}\right]}\right)\right]\\
 & \quad-\mathrm{E}\left[\log\left(\frac{1+P_{\mathrm{int},i}}{1+\mathrm{E}\left[P_{\mathrm{int},i}\right]}\right)\right],\end{align*}
where $P_{\mathrm{sig},i}$ and $P_{\mathrm{int},i}$ are defined
in (\ref{eq:signal-power}) and (\ref{eq:interference-power}). The
following theorem calculates $\mathrm{E}\left[P_{\mathrm{sig},i}\right]$
and $\mathrm{E}\left[P_{\mathrm{int},i}\right]$ for finite dimensional
systems.

\begin{thm}
\label{thm:Average-sig-int-power}Let $\mathcal{B}_{i}$'s be randomly
constructed and $D_{i}=\mathrm{E}_{\mathcal{B}_{i}}\left[D\left(\mathcal{B}_{i}\right)\right]$
for all $1\le i\le m$. For randomly chosen $A_{\mathrm{on}}$ and
$i\in A_{\mathrm{on}}$, if $1\le s\le L$\begin{equation}
\mathrm{E}\left[P_{\mathrm{sig},i}\right]=\gamma_{i}\rho\frac{L}{s}\left[\left(1-D_{i}\right)\left(1-\frac{s-1}{L}\right)+D_{i}\frac{s-1}{L\left(L-1\right)}\right],\label{eq:zf-average-signal}\end{equation}
 and \begin{equation}
\mathrm{E}\left[P_{\mathrm{int},i}\right]=\gamma_{i}\rho\frac{L}{s}\frac{s-1}{L-1}D_{i};\label{eq:zf-average-interference}\end{equation}
if $s>L$, $\mathrm{E}\left[P_{\mathrm{sig},i}\right]=0$. 
\end{thm}

The calculation of $\mathrm{E}\left[P_{\mathrm{sig},i}\right]$ and
$\mathrm{E}\left[P_{\mathrm{int},i}\right]$ relies on quantification
of $D_{i}$. In general, it is difficult to compute $D_{i}$ precisely.
Note that the upper bound in (\ref{eq:quantization-bds}) is derived
by evaluating the average performance of random codebooks (see \cite{Dai_05_Quantization_Grassmannian_manifold}
for details). We use its main order term to estimate $D_{i}$: \[
D_{i}\approx\frac{\Gamma\left(\frac{1}{L-1}\right)}{L-1}2^{-\frac{R_{i}}{L-1}}.\]

Define 

\begin{equation}
\mathcal{I}_{\mathrm{main},i}:=\log\left(1+\frac{\mathrm{E}\left[P_{\mathrm{sig},i}\right]}{1+\mathrm{E}\left[P_{\mathrm{int},i}\right]}\right).\label{eq:main-order-term}\end{equation}
It can be verified from Proposition \ref{pro:on-user-i-throughput}
that $\mathcal{I}_{i}=\mathcal{I}_{\mathrm{main},i}+o\left(1\right)$
and therefore $\mathcal{I}_{\mathrm{main},i}$ is the main order term
of $\mathcal{I}_{i}$.

Then the difference between asymptotic analysis and finite dimensional
systems analysis is clear. In the limit, $\frac{s-1}{L}\rightarrow\bar{s}$
and $\frac{R_{i}}{L-1}\rightarrow\bar{r}_{i}$. However, for finite
dimensional systems, simply substituting these asymptotic values into
(\ref{eq:zf-average-signal}-\ref{eq:main-order-term}) directly introduces
unpleasant error, especially when $L$ is small.  Therefore, to estimate
$\mathcal{I}_{i}$ ($\forall i\in A_{\mathrm{on}}$) for finite dimensional
systems, we have to rely on (\ref{eq:zf-average-signal}-\ref{eq:main-order-term}).

\subsection{A Scheme for Finite Dimensional Systems}

Given system parameters $L$, $m$, $\gamma_{i}$'s and $R_{i}$'s,
a practical scheme needs to calculate the appropriate $s$ and $A_{\mathrm{on}}$.
This process is described in the following.

For a given $s$, the set of $A_{\mathrm{on}}$ is decided as follows:
we first calculate $\mathcal{I}_{\mathrm{main},1},\cdots,\mathcal{I}_{\mathrm{main},m}$
according to (\ref{eq:main-order-term}) and then choose the $s$
users with the largest $\mathcal{I}_{\mathrm{main},i}$'s to turn
on; if there exists any ambiguity, random selection is employed to
resolve it. For example, if $\mathcal{I}_{\mathrm{main},1}>\mathcal{I}_{\mathrm{main},2}>\cdots>\mathcal{I}_{\mathrm{main},m}$,
the user $1,2,\cdots,s$ are turned on. If $\mathcal{I}_{\mathrm{main},1}=\mathcal{I}_{\mathrm{main},2}=\cdots=\mathcal{I}_{\mathrm{main},m}$,
the $s$ on-users are randomly selected from all the $m$ users. Note
again, $A_{\mathrm{on}}$ is independent of the channel realization.

The appropriate $s$ is chosen as follows. Let \[
\mathcal{I}_{\mathrm{main}}\left(s\right)=\underset{A_{\mathrm{on}}:\;\left|A_{\mathrm{on}}\right|=s}{\max}\;\sum_{i\in A_{\mathrm{on}}}\mathcal{I}_{\mathrm{main},i}.\]
Here, note that $\mathcal{I}_{\mathrm{main},i}$ is a function of
$s$. For a given broadcast system, we choose the number of on-users
to be \[
s_{\mathrm{main}}^{*}=\underset{1\le s\le L}{\arg\;\max}\;\mathcal{I}_{\mathrm{main}}\left(s\right).\]

Although the above procedure involves exhaustive search, the corresponding
complexity is actually low. First, the calculations are independent
of instantaneous channel realizations. Only system parameters $L$,
$m$, $\gamma_{i}$'s and $R_{i}$'s are needed. Provided that $\gamma_{i}$'s
change slowly, the base station does not need to recalculate $s_{\mathrm{main}}^{*}$
and $A_{\mathrm{on}}$ frequently. Second, $R_{i}=R_{j}$ in most
systems. For such systems and a given $s$, the $s$ on-users are
just simply the users with the largest $\gamma_{i}$'s. 

After calculating $s_{\mathrm{main}}^{*}$ and $A_{\mathrm{on}}$,
the base station broadcast $A_{\mathrm{on}}$ to all the users. For
each fading block, the system works as follows. 

\begin{itemize}
\item At the beginning of each fading block, the base station broadcasts
a single channel training sequence to help all the users estimate
their channel states $\mathbf{h}_{i}$'s. 
\item After estimating their $\mathbf{h}_{i}$'s, the on-users quantize
$\mathbf{h}_{i}$'s into $\mathbf{p}_{i}$'s according to (\ref{eq:quantization-fn})
and feed the corresponding indices to the base station. 
\item The base station then calculates the transmit beamforming vectors
$\mathbf{q}_{i}$'s according to (\ref{eq:zero-forcing-rule}), and
then transmits $\mathbf{q}_{i}X_{i}$'s. 
\end{itemize}
\begin{remrk}
[Fairness Scheduling]For systems with $\gamma_{i}\neq\gamma_{j}$
or $R_{i}\neq R_{j}$, there may be some users always turned off according
to the above scheme. Fairness scheduling is therefore needed to ensure
fairness of the system. There are many ways to perform fairness scheduling.
Since fairness is not the primary concern of this paper, we only give
an example as follows. Given $m$ users, the base station calculates
the corresponding $s_{\mathrm{main}}^{*}$ and $A_{\mathrm{on}}$,
and then turns on the users in $A_{\mathrm{on}}$ for the first fading
block. At the second fading block, the base station considers the
users who have not been turned on $\left\{ 1,\cdots,m\right\} \backslash A_{\mathrm{on}}$.
It calculates the corresponding $s_{\mathrm{main}}^{*}$ and $A_{\mathrm{on}}$,
and then turns on the users in the new $A_{\mathrm{on}}$. Proceed
this process until all users have been turned on once. Then start
a new scheduling cycle.
\end{remrk}

\subsection{Simulation Results}

\begin{figure}
\subfigure[$R_{\mathrm{fb}}=6$ Bits/Channel Realization]{\includegraphics[scale=0.6]{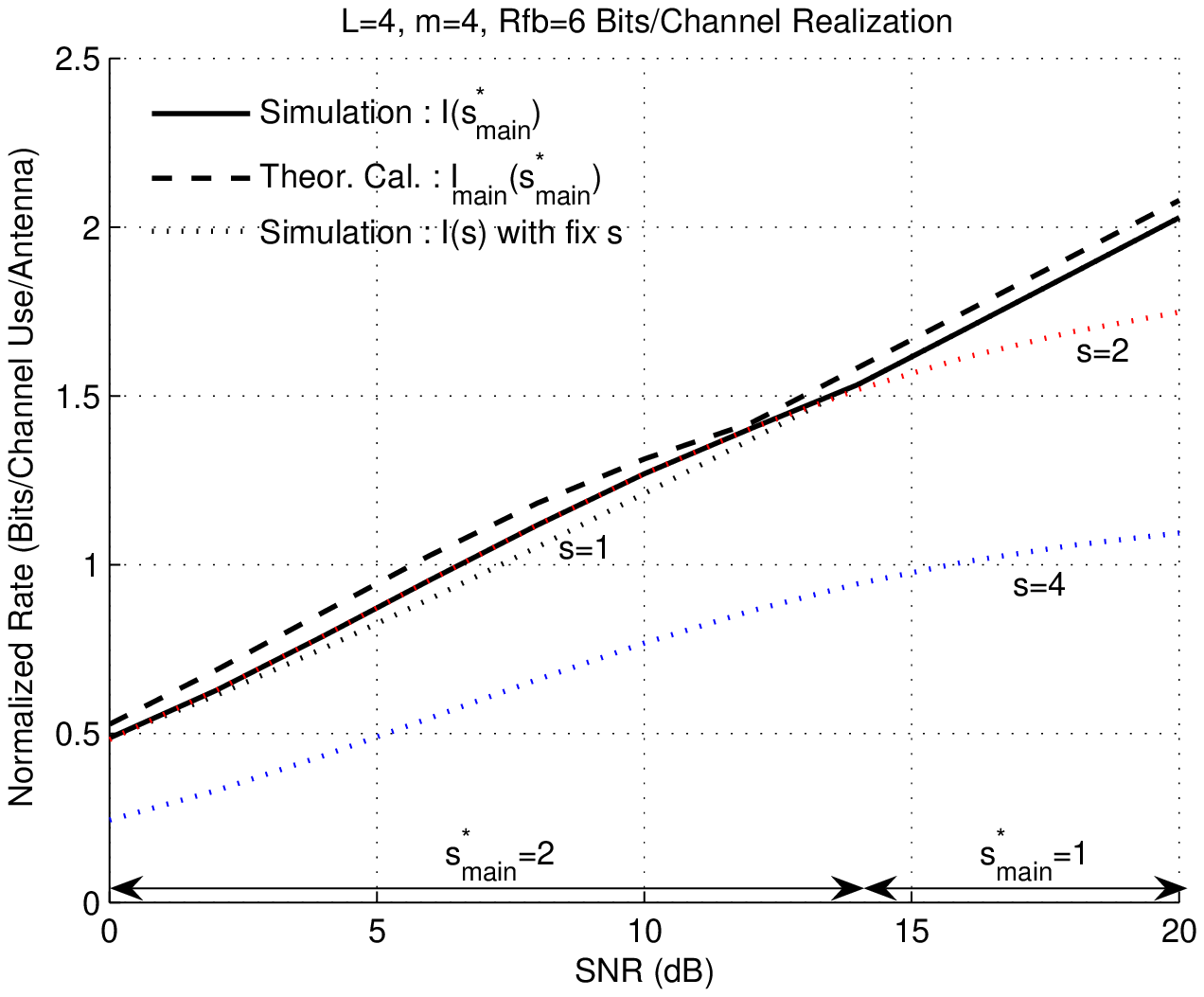}}

\subfigure[$R_{\mathrm{fb}}=12$ Bits/Channel Realization]{\includegraphics[scale=0.6]{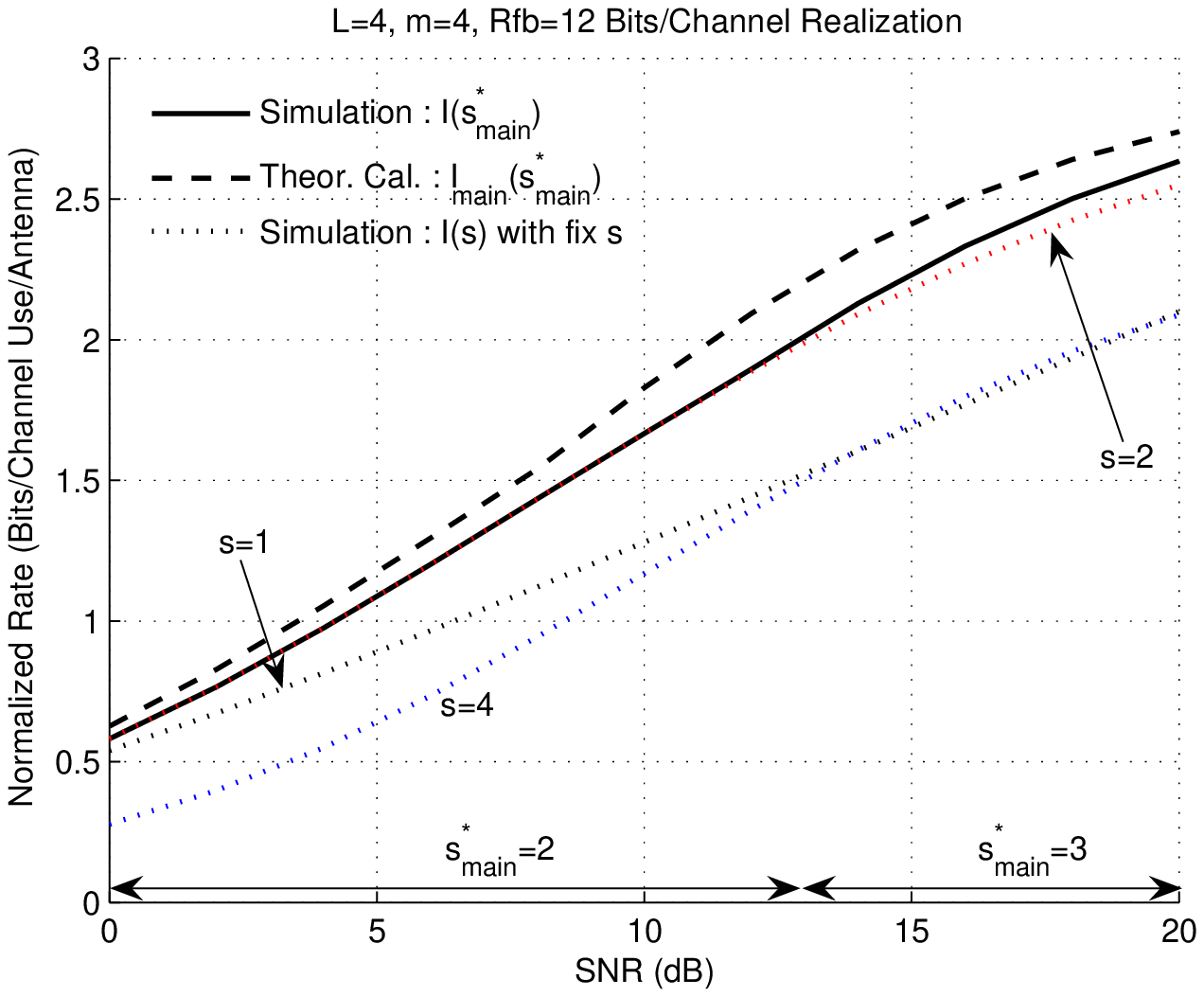}}

\caption{\label{cap:Rate-ZF}Total Throughput for Zero Forcing Beamforming}
\end{figure}

Fig. \ref{cap:Rate-ZF} gives the simulation results for the proposed
scheme using zero-forcing. In the simulations, $L=m=4$. For simplicity,
we assume that $\gamma_{1}=\gamma_{2}=\cdots=\gamma_{m}=1$ and $R_{1}=R_{2}=\cdots=R_{m}=R_{\mathrm{fb}}$.
With these assumptions, the $s$ on-users can be randomly chosen from
all the $m$ users. Without loss of generality, we assume that $A_{\mathrm{on}}\equiv\left\{ 1,\cdots,s\right\} $.
Let $\mathcal{I}\left(s\right)=\sum_{i\in A_{\mathrm{on}}}\mathcal{I}_{i}.$
In Fig. \ref{cap:Rate-ZF}, the solid lines are the simulations of
$\mathcal{I}\left(s_{\mathrm{main}}^{*}\right)$ while the dashed
lines are the theoretical calculation of $\mathcal{I}_{\mathrm{main}}\left(s_{\mathrm{main}}^{*}\right)$.
The simulation results show that the optimal $s$ is a function of
$\rho$ and $R_{\mathrm{fb}}$. For example, $s=1$ is optimal when
$\rho\in\left[15,20\right]$dB and $R_{\mathrm{fb}}=6$ bits, while
$s=3$ is optimal for the same SNR region as $R_{\mathrm{fb}}$ increases
to 12 bits. The reason behind it is that the interference introduced
by finite rate quantization is larger when $R_{\mathrm{fb}}$ is smaller:
when $R_{\mathrm{fb}}$ is small, the base station needs to turn off
some users to avoid strong interference as SNR gets very large. 

We also compare our scheme with the schemes where the number of on-users
is a presumed constant (independent of $\rho$ and $R_{\mathrm{fb}}$).
The throughput of schemes with presumed $s$ is presented in dotted
lines. From the simulation results, the throughput achieved by choosing
appropriate $s$ is always better than or equals to that with presumed
$s$. Specifically, compared to the scheme in \cite{Jindal_IT06sub_BC_Feedback}
where $s=L=4$ always, our scheme achieves a significant gain at high
SNR by turning off some users.

\section{\label{sec:Conclusion}Conclusion}

This paper considers heterogeneous broadcast systems with a relatively
small number of users. Asymptotic analysis where $L,m,s,R_{i}\rightarrow\infty$
linearly is employed to get insight into system design. Based on the
asymptotic analysis, we derive the asymptotically optimal feedback
strategy, propose a realistic on/off criterion, and quantify the spacial
efficiency. The key observation is that the number of on-users should
be appropriately chosen as a function of system parameters. Finally,
a practical scheme is developed for finite dimensional systems. Simulations
show that this scheme achieves a significant gain compared with previously
studied schemes with presumed number of on-users. 

\bibliographystyle{IEEEtran}
\bibliography{Bib/_BC_Feedback,Bib/_Jindal,Bib/_Liu_Dai,Bib/_Tse}

\end{document}